\documentclass[12pt]{article}
\usepackage{latexsym}
\usepackage[cp1250]{inputenc}

\title{Comment on "Dirac Quantization  of Pais-Uhlenbeck Fourth Order Oscillator"}
\author{Katarzyna Bolonek\thanks{supported by the grant 1 P03B 125 29 of the Polish Ministry of Science and  by The European Social Fund and Budget of 
State implemented under The Integrated Regional Operational Programme - Project GRRI-D .}\\ 
 Piotr  Kosi\'nski\thanks{supported by the grant 1 P03B 021 28 of the Polish Ministry of Science.} \\
Department of Theoretical Physics II \\
University of {\L}\'od\'z \\
Pomorska 149/153, 90 - 236 {\L}\'od\'z, Poland.}
\date{}

\begin{document}
\maketitle
\begin{abstract}
The structure of Pais-Uhlenbeck oscillator in the equal-frequency limit has been recently studied by Mannheim and Davidson
[Phys.Rev. A71 (2005), 042110]. It appears that taking this limit , as presented in the above paper,
 is quite subtle and the resulting structure of space of states - 
involved. In order to clarify the situation we present here the proper way of taking the 
equal-frequency limit, first under the assumption that the scalar product in the space of states is positive defined.
We discuss also the case of indefinite metric space of states. We show that, irrespective of the way the limit is defined, 
the limiting theory can be hardly viewed as satisfactory.

\end{abstract}

\newpage

Pais-Uhlenbeck quartic oscillator \cite{b1} is described by the Lagrangian 
\begin{eqnarray}
L=\frac{m}{2}\dot q^2-\frac{m\omega ^2}{2}q^2-\frac{m\lambda }{2}\ddot q^2   \label{w1}
\end{eqnarray}
Its behaviour depends on actual values of parameters $\omega $\ and $\lambda $. In what follows we keep $m$\ and $\omega $\ 
fixed while varying $\lambda $. The relevant equation of motion reads
\begin{eqnarray}
\lambda \left(\frac{d^2}{dt^2}+\omega ^2_1\right)\left(\frac{d^2}{dt^2}+\omega ^2_2\right)q(t)=0   \label{w2}
\end{eqnarray}
with
\begin{eqnarray}
\omega ^2_{1,2}\equiv \frac{1\pm \sqrt{1-4\lambda \omega ^2}}{2\lambda }   \label{w3}
\end{eqnarray}
For large $\lambda \;\;\; (\lambda >\frac{1}{4\omega ^2}) $\ both frequencies are complex. On the other hand, in the range 
$0<\lambda <\frac{1}{4\omega ^2}$\ they are real; in the limiting case $\lambda =\frac{1}{4\omega ^2}$\ there is a double 
degeneracy $\omega _1=\omega _2=\sqrt{2}\omega $. Finally, if $\lambda <0$, one frequency is real while the second one - 
purely imaginary. 

In order to quantize our theory one has to put it first in Hamiltonian form. This can be achieved within Ostrogradski formalism 
\cite{b2}, \cite{b3}. It is well known \cite{b4}, \cite{b5}, \cite{b6} that the Ostrogradski procedure is essentially a form 
of Dirac method for constrained theories. \\
In our case one finds \cite{b7}, \cite{b8} $\div $\ \cite{b10} the following canonical variables 
$$ q_1\equiv q, \;\;\; q_2\equiv \dot q $$
\begin{eqnarray}
&& \Pi _1\equiv \frac{\delta L}{\delta \dot q}=\frac{\partial L}{\partial \dot q}-\frac{d}{dt}\left(\frac{\partial L}
{\partial \ddot q}\right)=m(\dot q+\lambda \stackrel {\ldots}{q})    \nonumber \\
&& \Pi _2\equiv \frac{\delta L}{\delta \ddot q}=\frac{\partial L}{\partial \ddot q}=-m\lambda \ddot q   \label{w4}
\end{eqnarray}
together with the Hamiltonian
\begin{eqnarray}
H=\Pi _1q_2-\frac{1}{2m\lambda }\Pi _2^2+\frac{m\omega ^2}{2}q_1^2-\frac{m}{2}q_2^2   \label{w5}
\end{eqnarray}
Quantization can be now performed in the standard way by imposing the commutation rule
\begin{eqnarray}
[\hat q_i,\hat \Pi _j]=i\hbar \delta _{ij}  \label{w6}
\end{eqnarray}

Consider first the range $0<\lambda <\frac{1}{4\omega ^2}$; then $\omega _1^2>\omega _2^2>0$. To make the structure of the Hamiltonian transparent we 
perform the following canonical transformation (cf. Ref.\cite{b1})
\begin{eqnarray}
&& \hat q_1=\frac{1}{\sqrt{\lambda (\omega _1^2-\omega _2^2)}}(-\hat x_1+\hat x_2)   \nonumber \\ 
&& \hat q_2=\frac{1}{m\sqrt{\lambda (\omega _1^2-\omega _2^2)}}(\hat p_1+\hat p_2)  \label{w7} \\
&& \hat \Pi _1=\sqrt{\frac{\lambda }{\omega _1^2-\omega _2^2}}(\omega _2^2\hat p_1+\omega _1^2\hat p_2)   \nonumber \\
&& \hat \Pi _2=m\sqrt{\frac{\lambda }{\omega _1^2-\omega _2^2}}(-\omega _1^2\hat x_1+\omega _2^2\hat x_2)  \nonumber
\end{eqnarray}
Note that the above transformation becomes singular in the doubly degenerate limit $\lambda \rightarrow \frac{1}{4\omega ^2}$. In term of new variables 
the Hamiltonian takes particularly simple form
\begin{eqnarray}
\hat H=\left(\frac{\hat p_2^2}{2m}+\frac{m\omega _2^2}{2}\hat x_2^2\right)-\left(\frac{\hat p_1^2}{2m}+\frac{m\omega _1^2}{2}\hat x_1^2\right)  \label{w8}
\end{eqnarray}
The eigenvectors of $\hat H$\ are uniquely determined (up to a phase factor) by two nonnegative integers $n_1,n_2$\
\begin{eqnarray}
\hat H\mid n_1,n_2\rangle =\left(-\hbar \omega _1(n_1+\frac{1}{2})+\hbar \omega _2(n_2+\frac{1}{2})\right)\mid n_1,n_2\rangle  \label{w9}
\end{eqnarray}
The spectrum of $\hat H$\ is simple provided $\frac{\omega _1}{\omega _2}$\ is irrational; for rational $\frac{\omega _1}{\omega _2}$\ (superintegrable 
case) there is a degeneracy. \\
The wave functions in the coordinate representation read
\begin{eqnarray}
&&\langle x_1,x_2\mid n_1,n_2\rangle = \label{w10} \\
&&=N(n_1)N(n_2)^4\sqrt{\frac{m^2\omega _1\omega _2}{\hbar ^2}}H_{n_1}\left(x_1\sqrt{\frac{m\omega _1}{\hbar }}\right)
H_{n_2}\left(x_2\sqrt{\frac{m\omega _1}{\hbar }}\right)\cdot e^{-\frac{m}{2\hbar }(\omega _1x_1^2+\omega _2x_2^2)}  \nonumber
\end{eqnarray}
where $N(n)\equiv (\sqrt{\pi  }\cdot 2^n\cdot n!)^{-\frac{1}{2}}$. \\
The spectrum of $\hat H$, as given by eq.(\ref{w9}), is unbounded from below. One gets positive energy spectrum by admitting indefinite metric in the space 
of states. To this end we consider the space of states endowed with the positive-definite scalar product $(\cdot ,\cdot )$\ and define the "physical" scalar 
product with the help of metric operator $\eta $\
\begin{eqnarray}
\langle \Phi \mid \Psi \rangle \equiv (\Phi ,\eta \Psi ), \;\;\;\; \eta =\eta ^+=\eta ^{-1}  \label{w11}
\end{eqnarray}
Denoting by $"\star "$\ the hermitean conjugate with respect to the scalar product $\langle \cdot  \mid   \cdot  \rangle$\ one finds for 
any operator $\hat A$.
\begin{eqnarray}
\hat A^\star =\eta \hat A^+\eta    \label{w12}
\end{eqnarray}
Let $a_i,a^+_i$\ be the creation/anihilation operators constructed out of $\hat x_i,\hat p_i$,
\begin{eqnarray}
&& \hat x_i=i\sqrt{\frac{\hbar }{2m\omega _i}}\left(a_i-a^+_i\right)  \nonumber \\
&& \hat p_i=\sqrt{\frac{m\hbar \omega _i}{2}}\left(a_i+a^+_i\right)  \label{w13}
\end{eqnarray}
We define
\begin{eqnarray}
\eta =(-1)^{N_1}=e^{i\pi a^+_1a_1}   \label{w14}
\end{eqnarray}
Then
\begin{eqnarray}
\langle n_1,n_2\mid n'_1,n'_2\rangle =(-1)^{n_1}\delta _{n_1n'_1}\delta _{n_2n'_2};    \label{w15}
\end{eqnarray}
moreover, $\hat x^\star _i=\hat x_i,\; \hat p^\star _i=\hat p_i$\ imply
\begin{eqnarray}
\hat x^+_i=(-1)^i\hat x_i, \;\;\; \hat p^+_i=(-1)^i\hat p_i    \label{w16}
\end{eqnarray}
Defining (c.f. \cite {b1})
\begin{eqnarray}
&& \hat x'_1=\pm i\hat x_1, \;\;\; \hat x'_2=\hat x_2   \nonumber \\
&& \hat p'_1=\mp i\hat p_1, \;\;\;  \hat p'_2=\hat p_2   \label{w17}
\end{eqnarray}
we find that $\hat x'_i,\hat p'_i$\ are hermitean (with respect to "+" conjugation) and 
\begin{eqnarray}
\hat H=\left(\frac{\hat p'^2_1}{2m}+\frac{m\omega _1^2}{2}\hat x'^2_1\right)+\left(\frac{\hat p_2'^2}{2m}+\frac{m\omega _2^2}{2}\hat x'^2_2\right) 
\label{w18}
\end{eqnarray}
Therefore, the spectrum of $\hat H$\ is now positive definite
\begin{eqnarray}
E_{n_1,n_2}=\hbar \omega _1(n_1+\frac{1}{2})+\hbar \omega _2(n_2+\frac{1}{2})   \label{w19}
\end{eqnarray}
The "physical" subspace is spanned by the vectors $\mid 2n_1,n_2\rangle $. 

Let us now consider the degenerate case $\lambda =\frac{1}{4\omega ^2}$. To reveal the structure of the Hamiltonian we define new variables $\hat Q_i, 
\hat P_i$\ by \cite{b1} 
\begin{eqnarray}
&& \hat q_1=\frac{\hat Q_1}{2\sqrt{2}}+\frac{\hat P_2}{m\omega }  \nonumber \\
&& \hat q_2=\frac{\omega \hat Q_2}{2}+\frac{\sqrt{2}\hat P_1}{m}  \label{w20} \\
&& \hat \Pi _1=\frac{\hat P_1}{\sqrt{2}}-\frac{3}{4}m\omega \hat Q_2   \nonumber \\
&& \hat \Pi _2=\frac{\hat P_2}{2\omega }-\frac{3}{4\sqrt{2}}m\hat Q_1  \nonumber
\end{eqnarray}
Eq.(\ref{w20}) defines a $\lambda $\- independent canonical transformation. In terms of new variables the Hamiltonian takes a particularly simple form
\begin{eqnarray}
\hat H=\sqrt{2}\omega (\hat Q_1\hat P_2-\hat Q_2\hat P_1)-\frac{m\omega ^2}{2}(\hat Q_1^2+\hat Q_2^2)   \label{w21}
\end{eqnarray}
Again, it is a sum of two commutating pieces: the first proportional to the angular momentum operator while the second represents the length of 
$\hat {\vec Q}$\ squared. Therefore, the energy is the sum of discrete and continuous parts and depends on the combination of two quantum numbers \cite {b1}. 

In what follows we find it convenient to use the momentum representation: $\hat P_i\rightarrow P_i, \;\; \hat Q_i\rightarrow i\hbar \frac{\partial }
{\partial P_i}$. Let us define the polar coordinates in momentum space by $P_1=P\cos \Theta , \;\; P_2=P\sin \Theta $. Then
\begin{eqnarray}
\hat H=-i\sqrt{2}\omega \hbar \frac{\partial }{\partial \Theta }+\frac{m\omega ^2\hbar ^2}{2}\left(\frac{\partial ^2}{\partial P^2}+\frac{1}{P}
\frac{\partial }{\partial P}+\frac{1}{P^2}\frac{\partial ^2}{\partial \Theta ^2}\right);  \label{w22}
\end{eqnarray}
the spectrum and normalized eigenfunctions read
\begin{eqnarray}
&&E_{n,k}=\omega \hbar \left(\sqrt{2}n-\frac{m\omega \hbar k^2}{2}\right) \label{w23} \\
&& \Psi _{n,k}(\vec P)=\sqrt{\frac{k}{2\pi  }}J_n(kP)e^{in\Theta }  \nonumber
\end{eqnarray}

We shall now consider the limit $\lambda \rightarrow \frac{1}{4\omega ^2}$. This is slightly subtle due to the fact that the spectrum of the
Hamiltonian changes in this limit from discrete into continuous one. Let us put
\begin{eqnarray}
1-4\lambda \omega ^2\equiv \varepsilon ^2, \;\;\; \varepsilon \rightarrow 0^+  \label{w24}
\end{eqnarray}
Then
\begin{eqnarray}
\omega _{1,2}\simeq \sqrt{2}\omega \left(1\pm \frac{\varepsilon }{2}\right)  \label{w25}
\end{eqnarray}
The energy spectrum (\ref{w9}) can be rewritten as
\begin{eqnarray}
E_{n_1,n_2}=\sqrt{2}\omega \hbar (n_2-n_1)-\frac{\sqrt{2}}{2}\omega \hbar \varepsilon (n_1+n_2+1)     \label{w26}
\end{eqnarray}
In the limit $\varepsilon \rightarrow 0^+$\ the energy seems to be given by the formula  \cite{b11}
\begin{eqnarray}
E_{n_1,n_2}=\sqrt{2}\omega \hbar (n_2-n_1)     \label{w27}
\end{eqnarray}
This is, however, not the case. We should take into account that the limiting Hamiltonian has a continuous spectrum given by eq.(\ref{w24}). Therefore, 
the proper way of taking the limit is to let $n_1,n_2\rightarrow \infty $\ in such a way that
\begin{eqnarray}
&& n=n_2-n_1   \nonumber  \\
&& \varepsilon (n_1+n_2)=\frac{m\omega \hbar k^2}{\sqrt{2}}    \label{w28}
\end{eqnarray}
are fixed. \\
We shall show that, indeed, by considering this limiting procedure one recovers the wave functions (\ref{w23}) of degenerate Hamiltonian.  

Let us note that we cannot take the equal frequency limit directly. This is due to the fact that the very coordinate representation becomes singular in 
this limit as is clearly seen from eq.(\ref{w7}). On the other hand, the momentum representation based on $\hat P_1, \hat P_2$\ (cf. eqs.(\ref{w22}), 
(\ref{w23})) is always well-defined. Therefore, the first step will be to rewrite our wavefunctions (\ref{w10}) in momentum representation. To this end 
we write $\hat P_1$\ and $\hat P_2$\ in terms of $\hat x_i's$\ and $\hat p_i's$. By virtue of eqs.(\ref{w7}) and (\ref{w20}) we find
\begin{eqnarray}
&& \hat P_1=\frac{b_2\hat p_1+b_1\hat p_2}{m\sqrt{\lambda (\omega _1^2-\omega _2^2)}}=\frac{-i\hbar }{m\sqrt{\lambda (\omega _1^2-\omega _2^2)}}
\left(b_2\frac{\partial }{\partial x_1}+b_1\frac{\partial }{\partial x_2}\right)  \nonumber \\
&& \hat P_2=\frac{\sqrt{2}\omega (-b_1\hat x_1+b_2\hat x_2)}{\sqrt{\lambda (\omega _1^2-\omega _2^2)}}=\frac{\sqrt{2}\omega }{\sqrt{\lambda (\omega _1^2
-\omega _2^2)}}\left(-b_1x_1+b_2x_2\right)  \label{w29}  \\
&& b_{1,2}\equiv \frac{m}{2\sqrt{2}}\left(\frac{3}{2}+\lambda \omega _{1,2}^2\right)   \nonumber
\end{eqnarray}
It is now straightforward to find the relevant transition functions $\langle x_1,x_2\mid P_1,P_2 \rangle$\ by solving the corresponding eigenvalue equations
\begin{eqnarray}
\hat P_i\langle x_1,x_2\mid P_1,P_2\rangle =P_i\langle x_1,x_2\mid P_1,P_2\rangle   \label{w30}
\end{eqnarray}
Using eqs.(\ref{w29}) one obtains
\begin{eqnarray}
&&\langle x_1,x_2\mid P_1,P_2 \rangle = \label{w31} \\
&& =\sqrt{\frac{m\lambda (\omega _1^2-\omega _2^2)}{2\sqrt{2}\pi \hbar \omega }}\delta \left(-b_1x_1+b_2x_2-
\frac{\sqrt{\lambda (\omega _1^2-\omega _2^2)}P_2}{\sqrt{2}\omega }\right)e^{
\frac{im\sqrt{\lambda (\omega _1^2-\omega _2^2)}}{\hbar (b_1^2+b_2^2)}P_1(b_2x_1+b_1x_2)}  \nonumber
\end{eqnarray}
This allows us to pass to the momentum representation
\begin{eqnarray}
\langle P_1,P_2\mid n_1,n_2\rangle =\int dx_1dx_2\langle P_1,P_2\mid x_1,x_2\rangle \langle x_1,x_2\mid n_1,n_2 \rangle  \label{w32}
\end{eqnarray}
Doing one integration with the help of delta function one arrives at the following rather complicated expression:
\begin{eqnarray}
&& \langle P_1,P_2\mid n_1,n_2\rangle = \frac{N(n_1)N(n_2)}{b_1^2+b_2^2}\sqrt[4]{\frac{m^2\omega _1\omega _2}{\hbar ^2}}\sqrt{\frac{m\lambda (\omega _1^2
-\omega _2^2)}{2\sqrt{2}\pi \hbar \omega }}\cdot  \nonumber \\
&& \cdot e^{\frac{m}{2\hbar }\left[\frac{\lambda (\omega _1^2-\omega _2^2)}{\omega _1b_2^2+\omega _2b_1^2}\left(\frac{b_1b_2(\omega _2-\omega _1)}
{\sqrt{2}\omega (b_1^2+b_2^2)}P_2+iP_1\right)^2-\frac{\lambda (\omega _1^2-\omega _2^2)}{2\omega ^2}\frac{(\omega _1b_1^2+\omega _2b_2^2)}
{(b_1^2+b_2^2)^2}P_2^2\right]}\cdot   \label{w33}  \\
&& \cdot \int\limits_{-\infty }^{+\infty }dvH_{n_1}\left(\sqrt{\frac{m\omega _1}{\hbar }}\left(\frac{b_2v-b_1\sqrt{\frac{\lambda (\omega _1^2-\omega _2^2)}
{\sqrt{2}\omega }}P_2}{b_1^2+b_2^2}\right)\right)H_{n_2}\left(\sqrt{\frac{m\omega _2}{\hbar }}\left(\frac{b_1v+\frac{\sqrt{\lambda (\omega _1^2-\omega _2^2)}}
{\sqrt{2}\omega }P_2}{b_1^2+b_2^2}\right)\right)\cdot   \nonumber  \\
&& \cdot e^{\frac{-m}{2\hbar }\left[\frac{\sqrt{\omega _1b_2^2+\omega _2b_1^2}}{b_1^2+b_2^2}v+\sqrt{\frac{\lambda (\omega _1^2-\omega _2^2)}
{\omega _1b_2^2+\omega _2b_1^2}}\left(\frac{b_1b_2(\omega _2-\omega _1)}{\sqrt{2}\omega (b_1^2+b_2^2)}P_2+iP_1\right)\right]^2}  \nonumber
\end{eqnarray}
In principle, the last integral could be also taken. However, this is not necessary. We can take the limit $\varepsilon \rightarrow 0^+$\ directly in eq.
(\ref{w33}). In this limit
\begin{eqnarray}
b_{1,2}\simeq \frac{m}{\sqrt{2}}\left(1\pm \frac{\varepsilon }{4}\right)   \label{w34}
\end{eqnarray}
Keeping dominant terms in eq.(\ref{w33}) one finds
\begin{eqnarray}
&& \langle P_1,P_2\mid n_1,n_2\rangle\simeq \frac{N(n_1)N(n_2)\sqrt{\varepsilon }}{\sqrt{\pi }\hbar }\cdot   \label{w35}  \\
&& \cdot \int\limits_{-\infty }^{+\infty }dyH_{n_1}\left(y-\sqrt{\frac{\varepsilon }{2\sqrt{2}m\hbar \omega }}(P_2+P_1)\right)H_{n_2}\left(y-\sqrt\frac
{\varepsilon }{2\sqrt{2}m\hbar \omega }(-P_2+iP_1)\right)e^{-y^2} \nonumber
\end{eqnarray}
The last integral is taken explicitly yielding \cite{b12}
\begin{eqnarray}
&& \langle P_1,P_2\mid n_1,n_2\rangle\simeq  \label{w36} \\
&& \simeq  \frac{\sqrt{\varepsilon }}{\sqrt{\sqrt{2}m\hbar \omega }}\left(\frac{\varepsilon }{2\sqrt{2}m\hbar \omega }\right)^
{\frac{n}{2}}(P_2-iP_1)^nn_1!2^{n_2}N(n_1)N(n_2)L^n_{n_1}\left(\frac{\varepsilon (P_1^2+P_2^2)}{\sqrt{2}m\omega \hbar }\right)  \nonumber
\end{eqnarray}
(for definiteness we have assumed here $n_1\geq n_2$; the opposite case goes along the same way). Now, it is easy to take the limit $\varepsilon 
\rightarrow 0, \;\; n_{1,2}\rightarrow \infty $\ with $n\equiv n_2-n_1$\ and $\varepsilon (n_1+n_2)=\frac{m\omega \hbar k^2}{\sqrt{2}}$\ fixed. Using 
Stirling formula and the asymptotic form of Laguerre polynomials \cite{b12} we arrive finally at the following result
\begin{eqnarray}
\langle P_1,P_2\mid n_1,n_2\rangle_{n_1,n_2\rightarrow \infty  \varepsilon \rightarrow 0}\simeq (-i)^n\sqrt{\frac{\sqrt{2}\varepsilon }{m\hbar \omega 
k}}\left(\sqrt\frac{k}{2\pi }J_n(kP)\right)e^{in\Theta }  \label{w37}
\end{eqnarray}
By comparying eq.(\ref{w36}) and (\ref{w23}) we conclude that by taking the limit of equal frequencies in the way prescribed above we recover the 
wavefunctions of degenerate Hamiltonian. To complete the arguments let us only note that the additional $\sqrt{\varepsilon }$\ factor comes from the fact 
that the discrete eigenfunctions have unit norm while the norm of limiting one is infinite. 

Let us now consider the question whether the double-frequency case can be quantized 
in the way which yields positive-definite Hamiltonian at the expense 
of having indefinite metric in the space of states. More precisely, the problem is the following: for differening frequencies we choose the 
positive-energy quantization scheme; is it possible to perform the equal-frequency limit keeping the energy positive?

In order to answer this question one has to adopt some definition of taking the equal-frequency limit. The most reasonable way is to keep the 
initial operators $\hat q_i,\hat \Pi _i$\  as given once forever linear operators acting in some linear space of states and varying the 
Hamiltonian by varying some of its parameters ($\lambda $\ in our case). For $\varepsilon >0$\ one can define, via eqs.(\ref {w7}) and (\ref {w16}), 
the hermitean conjugation $"+"$\ for $q'_is$\ and $\Pi '_is$\ (and, hence, convert the space of states into usual Hilbert space) in such a way that the 
Hamiltonian (\ref {w5}) becomes a positive-definite operator. The definition of hermitean conjugation $"+"$\ depends, of course, on $\varepsilon $. 
It is easy to compute, using (\ref {w7}) and (\ref {w16}), that, for example
\begin{eqnarray}
\hat q^+_1\simeq \frac{1}{\varepsilon ^2}\left(\hat q_1-\frac{2}{m}\hat \Pi _2\right)     \label {w38}
\end{eqnarray}
Therefore, the reasonable conjugation rule cannot be imposed in the limiting case. One can support this conclusion by considering the operators $a$\ 
and $b$\ defined by eqs. (33) of Ref. \cite {b8}. In fact, it is easy to check that a diverges if the limit is taken in the way prescribed above; 
this is not in contradiction with the fact that their commutators and the Hamiltonian (eqs. (35), (40) of Ref \cite {b8}) are well-behaving 
in the limit $\varepsilon \rightarrow 0$.

However, one can argue that our prescription of taking the $\varepsilon \rightarrow 0$\ limit is not the only possible. In fact, one can argue 
that it is sufficient to get the regular limiting commutation rules and the Hamiltonian. This seems to be the strategy adopted by the authors 
of Ref. \cite {b8}.

Let us consider in some detail  their construction. The limiting commutation rules and the Hamiltonian read
\begin{eqnarray}
&& [a,a^\star ]=0, \;\;\; [b,b^\star ]=0, \;\;\; [b,a^\star ]=\mu , \;\;\; [a,b^\star ]=\mu , \;\;\; [a,b]=0  \nonumber \\
&& H=\frac{\omega }{\mu }(2b^\star b+a^\star b+b^\star a) +const  \label {w39}
\end{eqnarray}
By defining
\begin{eqnarray}
A_1=\frac{1}{\sqrt{2\mu }}\left(a+b\right), \;\;\; A_2=\frac{1}{\sqrt{2\mu }}\left(a-b\right)   \label{w40}
\end{eqnarray}
one obtains
\begin{eqnarray}
&& [A_i,A_j]=0, \;\;\; [A^\star _i,A^\star _j]=0,  \;\;\;[A_i,A^\star _j]=0, i\neq j, \;\;\; [A_1,A^\star _1]=1,  \;\;\;\;  [A_2,A^\star _2]=-1   \nonumber \\
&& H=\omega (2A_1^\star A_1-A_1^\star A_2-A_2^\star A_1)   \label {w41}
\end{eqnarray}
Then we obtain the indefinite Fock space for two degrees of freedom. Define new metric operator
\begin{eqnarray}
\tau \equiv e^{i\pi A_2^+A_2}  \label {w42}
\end{eqnarray}
and new conjugation
\begin{eqnarray}
B^+\equiv \tau B^\star \tau   \label {w43}
\end{eqnarray}
Then $A^+_1=A_1^\star , \; A_2^+=-A_2^\star $\ and one obtains standard commutation rules together with Fock representation with positive metric 
and $"+"$\ playing the role of hermitean conjugation.

Now, $H$\ can be viewed as acting in standard Hilbert space,
\begin{eqnarray}
H=\omega (2A_1^+ A_1-A_1^+A_2+A_2^+A_1)   \label {w44}
\end{eqnarray}
We see that $H$\ is not normal
\begin{eqnarray}
[H,H^+]\neq 0    \label {w45}
\end{eqnarray}
so it cannot be diagonalized. Moreover,,
\begin{eqnarray}
[H,N]=0   \label {w46}
\end{eqnarray}
where $N\equiv A_1^+A_1+A_2^+A_2$\ is the total number operator. Therefore, we can consider $H$\ as acting separately in each finitedimensional 
subspace of fixed eigenvalue of $N$. It is represented as a matrix in any such subspace and we want to determine its Jordan form. To this end 
consider the subspace spanned by $\mid n_1,n_2 \rangle, \;n_1+n_2=n$; then
\begin{eqnarray}
&& (H-\omega n)^{n+1}\mid n_1,n_2\rangle=(H-\omega N)^{n+1}\mid n_1,n_2\rangle=  \nonumber \\
&& =\omega ^{n+1}(A_1^+A_1-A_2^+A_2-A_1^+A_2+A_2^+A_1)^{n+1}
\mid n_1n_2\rangle=  \nonumber \\
&& =\omega ^{n+1}((A_1^++A_2^+)(A_1-A_2))^{n+1}\mid n_1,n_2\rangle =  \nonumber \\
&& =\omega ^{n+1}(A_1^++A_2^+)^{n+1}(A_1-A_2)^{n+1}
\mid n_1,n_2\rangle =0  \label {w47}
\end{eqnarray}
 because $[A^+_1+A^+_2,\;A_1-A_2]=0$. Therefore, in any subspace corresponding to the eigenvalue $n$\ of $N$\ $H$\ takes the form 
of single Jordan block; any such subspace contains exactly one eigenvector corresponding to the eigenvalue $n\omega $. In order to construct it 
we note that
\begin{eqnarray}
[H,A^+_1+A^+_2]=\omega (A^+_1+A^+_2)   \label {w48}
\end{eqnarray}
Therefore
\begin{eqnarray}
\mid n\rangle=(A^+_1+A^+_2)^n\mid 0,0\rangle     \label {w49}
\end{eqnarray}
is the eigenvector of $H$\ corresponding to the eigenvalue $n\omega $\ of $H$. Now, using
\begin{eqnarray}
\mid n\rangle=\sum\limits_{k=0}^n{n\choose k}{\sqrt{k!}\sqrt{(n-k)!}\mid k,n-k\rangle}  \label {w50}
\end{eqnarray}
one finds
\begin{eqnarray}
\langle n\mid n\rangle=n!\sum\limits_{k=0}^n{n\choose k}{(-1)^{n-k}}=0   \label {w51}
\end{eqnarray}
We see that all eigenvectors of $H$\ have zero norm. However, it is reasonable to assume that the physical space of states is spanned by 
the eigenvectors of $H$. This is a subspace of zero norm so it can hardly be viewed as physical one.

Concluding, we see no way to get the positive-energy quantized degenerate PU theory, at least by limiting procedure.

\end{document}